\documentclass[prl,twocolumn,pdftex]{revtex4-1}
\usepackage{graphicx}
\usepackage{amsmath}
\usepackage{graphicx}
\usepackage{amsmath}

\begin{document}
\title{Charge conservation in RHIC and contributiuons to local parity violation observables}
\author{S\"{o}ren Schlichting and Scott Pratt}
\affiliation{Department of Physics and Astronomy and National Superconducting Cyclotron Laboratory,
Michigan State University\\
East Lansing, Michigan 48824}
\date{\today}
\begin{abstract}
Relativistic heavy ion collisions provide laboratory environments from which one can study the creation of a novel state of matter, the quark gluon plasma. The existence of such a state is postulated to alter the mechanism and evolution of charge production, which then becomes manifest in charge correlations. We study the separation of balancing charges at kinetic freeze-out by analyzing recent result on balancing charge correlations for Au+Au collisions at $\sqrt{s_{NN}}=200~\mbox{GeV}$. We find that balancing charges are emitted from significantly smaller regions in central collisions compared to peripheral collisions. The results indicate that charge diffusion is small and that the centrality dependence points to a change of the production mechanism. In addition we calculate the contributions from charge-balance correlations to STAR's local parity violation observable. We find that local charge conservation, when combined with elliptic flow, explains much of STAR's measurement.
\end{abstract}

\maketitle

\section{I. INTRODUCTION}
\noindent In relativistic heavy ion collisions a large fraction of the detected charge is produced throughout the time evolution of the system. In particular for Au+Au collisions at $\sqrt{s_{NN}}=200~\mbox{GeV}$ the charged particle multiplicity is initially $N_{ch}=158$, whereas in the final state $dN_{ch}/d\eta$ is observed to be many hundreds of particles per unit rapidity, over multiple units of rapidity, with the precise values dependent on the impact parameter \cite{STAR:2009_Mult}. Despite the necessity of a correct treatment of the charge production inherent to relativistic heavy ion collisions, different reaction models employ very different production mechanisms. In event generators like HIJING \cite{HIJING} and partonic/hadronic cascades such as URQMD \cite{URQMD} the charges are created early by the fragmentation of gluonic strings. In contrast, hydrodynamic models have the inherent assumption that a large fraction of the charge is produced at a later time at the deconfinement phase transition, i.e. when hadrons appear in the collision. Possible scenarios for the late stage production of charge include formation of hadrons from gluons, conversion of the nonperturbative vacuum energy into particles as well as quark production induced by the hadronization of quarks. In this context the late stage production of quarks has been proposed as a signature of the existence and lifetime of the quark gluon plasma \cite{Bass:2000_BalanceFunction}. In principle, charge production could occur at any time before the chemical freeze-out of the system. The problem is therefore related to the dynamics of the non-equilibrium phase, the quark-gluon plasma and finally the deconfinement phase transition. However, microscopic approaches from first principles are still in their infancy and can not yet provide further insight with regard to charge production mechanisms. It is therefore of great importance to explore the experimental possibilities to obtain valuable information. In the context of the search for late stage production of charges it has been proposed to investigate balancing charge correlations \cite{Bass:2000_BalanceFunction}. The idea is based on the principle that charge is created in balancing pairs that originate from the same point in space-time. In the subsequent rescattering of the charge carrier, which in principle can be hadronic or partonic, the balancing partners are then spread out within some finite distance of each other. However the motion is in general highly collective, i.e. the collision rates are high compared to the rate of expansion, so that the original correlation in space-time transforms into a correlation in momentum space in the emission profile at freeze-out. In a transport model the motion of the balancing partners could be separated into a collective mode, due the collective expansion of the system, and a diffusive relative mode, due to the collisions with other particles. The kinetic freeze-out configuration is then described by the single particle distributions with the additional constraint of having balancing partners emitted within a certain distance of each others, related to the diffusion constant and the available timescale for diffusion \cite{Bass:2000_BalanceFunction}. If charge is created late in the collision and diffusion is small the balancing partners are expected to be emitted within a small range and the correlation is strongest. The study of charge-balance correlations hence gives insight into the production and diffusion of charge.

The observable that is employed to investigate these correlations is the charge balance function, which in the most general form is given by
\begin{eqnarray}
 B(p_1,p_2)&=&\frac{N_{+-}(p_2|p_1)-N_{++}(p_2|p_1)}{dM/dp_1}\nonumber \\
&&+\frac{N_{-+}(p_2|p_1)-N_{--}(p_2|p_1)}{dM/dp_1} \;.
\label{eq:general_bf}
\end{eqnarray}
Here $dM/dp_1$ denotes the differential charged particle multiplicity and $N_{+-}(p_2|p_1)$ is the number of pairs with a positive charge emitted with momentum $p_1$ and a negative charge emitted with momentum $p_2$ and accordingly for $N_{++}$, $N_{-+}$ and $N_{--}$. The balance function is constructed in such a way that the like-sign subtraction statistically isolates the balancing partner. If we assume that all other correlations are charge insensitive, i.e. identical for same-sign and like-sign pairs, then these correlations are removed from the balance function by subtracting the like-sign pairs from the opposite-sign pairs. In reality the additional sources of correlations are charge sensitive. In particular it has been shown that for small relative momenta the balance function is sensitive to correlations induced by the final-state interactions between the two particles due to Coulomb interactions and identical particle symmetrization \cite{Cheng:2004_balance_models}. One needs to account for such correlations whenever $p_1 \approx p_2$ in Eq. (\ref{eq:general_bf}) above.\\
The balance function in (\ref{eq:general_bf}) is a six-dimensional function of the particles momenta. In the context of studies of the separation of balancing charges it is sensible to reduce the discussion to measures of the differences of the momenta, $p_1-p_2$. In particular we will focus on the charge balance function in relative pseudorapidity $\Delta \eta$ and relative azimuthal angle $\Delta \phi$, which are given by
\begin{eqnarray}
B(\Delta\eta)=\frac{1}{M}\int dp_1 \frac{dM}{dp_1} dp_2  B(p_1|p_2)  \delta(\Delta\eta-|\eta_1-\eta_2|) \;,
\end{eqnarray}
and accordingly for $\Delta \phi$. Balance functions of this form have been discussed in \cite{Bass:2000_BalanceFunction,Cheng:2004_balance_models} and been measured by the STAR collaboration \cite{STAR:2003_BalanceFunctions,STAR:2010_BalanceFunctions}. The experimental results have shown that balancing charges are in general highly correlated at freeze-out. The observed correlations are stronger for more central collisions compared to more peripheral collisions and remarkably it was found that for central collisions the observed charge-balance correlations are consistent with thermal emission of balancing pairs from the same collective velocity, i.e. perfectly local charge conservation at freeze-out \cite{Cheng:2004_balance_models,STAR:2010_BalanceFunctions}. However it could not be distinguished, how much of the observed narrowing of the balance function is due to a change in temperature and flow, as opposed to a smaller separation of balancing charges at freeze-out in coordinate space.\\
In this paper we extend previous studies by performing a systematic analysis of the locality of charge conservation at break up based on the most recent STAR measurement of charge balance functions \cite{STAR:2010_BalanceFunctions}. We modify STAR's single particle blast-wave model \cite{STAR:2005_v2} to account for local charge conservation at freeze-out. In this model particles are emitted according to the blast-wave description with the additional constraint of local charge balance within a finite range of longitudinal rapidity $\sigma_\eta$ and azimuthal angle $\sigma_\phi$. We then extract the separation of balancing charges by adjusting $\sigma_\eta$ and $\sigma_\phi$ to reproduce the observed charge balance functions in relative pseudo-rapidity $\Delta \eta$ and relative azimuthal angle $\Delta \phi$. Since the blast-wave parameters were adjusted by STAR to match spectra and elliptic flow for each centrality bin, this allows us to distinguish between effects that are caused by a change in the freeze-out temperature and the collective flow versus effects that are due to the spatial separation of balancing charges, which is only affected by the production mechanism and the subsequent charge diffusion.\\
In addition, we show how charge-balance correlations contribute to STAR's local parity violation observable \cite{STAR:2009_Parity} and use the same model to perform a detailed analysis of these ``background'' contributions. This paper is organized as follows: The modified blast-wave model is described in detail in Sec. II, results on charge separation are presented in Sec. III. In Sec. IV we give an introduction to the phenomenon of local parity violation in relativistic heavy ion collisions and show how the fluctuations of parity-odd observables measured at STAR are affected by charge-balance correlations. The findings are summarized in Sec. V.

\section{II. FREEZE-OUT MODEL}
\begin{figure}[hpt]
\centerline{
	\includegraphics[width=0.4\textwidth]{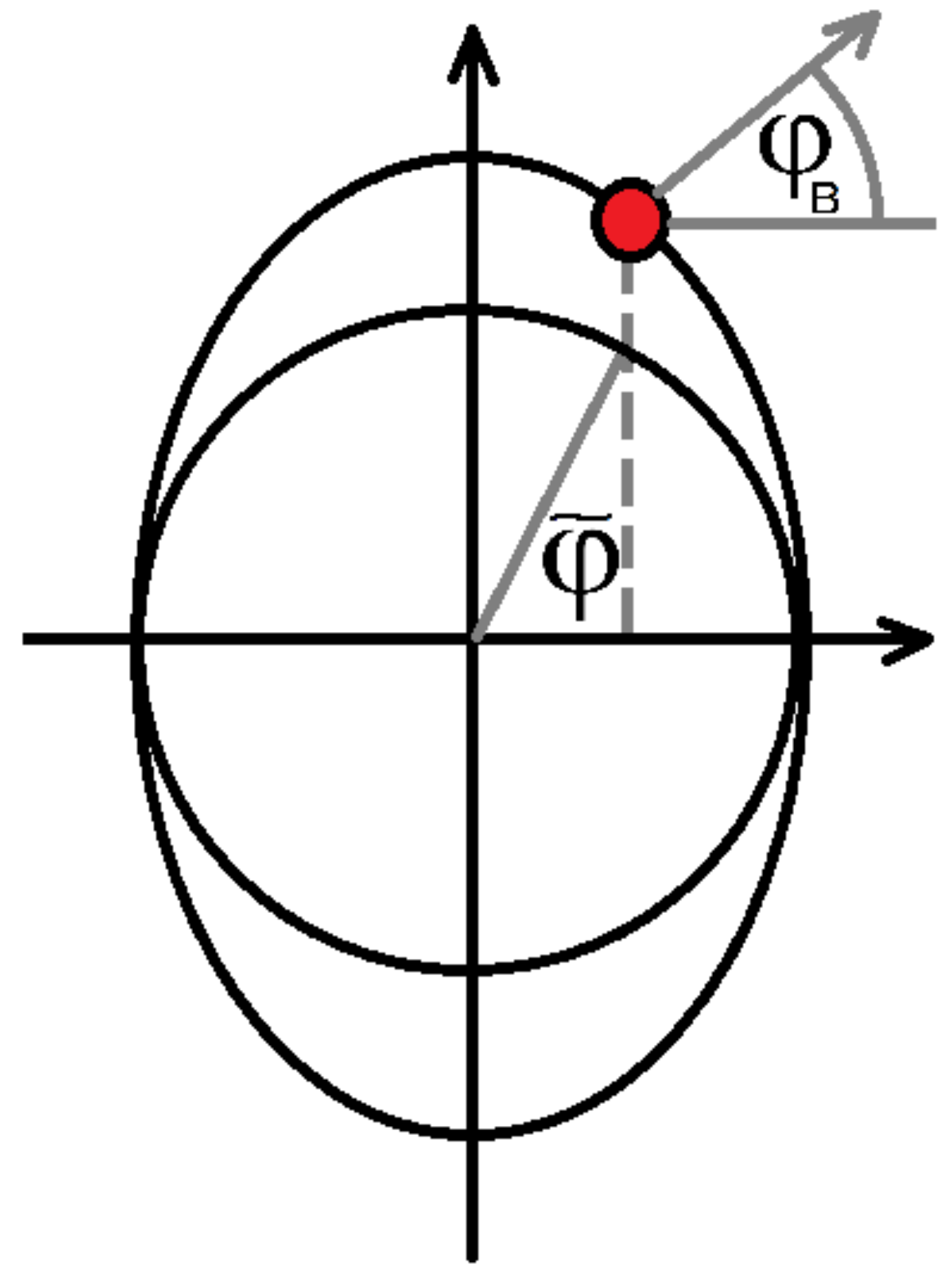}
}
\caption[Visualization of the blast-wave parameterization]%
	{\label{fig:BlastWave}
	Visualization of the blast-wave parameterization in the transverse plane. At a point $(x,y)$ in the transverse plane the associated collective velocity is perpendicular to the surface of the ellipse. According to (\ref{eq:BlastWave}) the magnitude of the transverse collective rapidity increases linearly with the reduced radius and has a second order harmonic component w.r.t the boost angle $\phi_B$. The relative distribution of the balancing charges (\ref{eq:Gauss}) in the transverse plane is a Gaussian in $\Delta\tilde{\phi}$.
}
\end{figure}

In common parameterizations of the kinetic freeze-out configuration, the system is described by a local thermal equilibrium,
\begin{eqnarray}
\frac{dN}{d^3p~d^4x}\propto \exp\left(-\frac{p^{\mu} u_{\mu}(x)}{T_{kin}}\right)~\chi_S(x),
\end{eqnarray}
where $\chi_S(x)$ is the characteristic function of the freeze-out surface $S$. The system is then completely described by the collective velocity profile $u_{\mu}(x)$, the kinetic freeze-out temperature $T_{kin}$ and the freeze-out surface $S$, which is usually chosen to be some volume at a fixed time in the lab frame. The conventional way to apply such a blast-wave model is then to choose a point on the freeze-out surface, create a particle with momentum according to the kinetic freeze-out temperature and the collective velocity, calculate the observable of interest and then repeat for several particles until sufficient statistics is achieved. The chemical composition is either governed by some parameterization of the chemical freeze-out configuration or enforced to match experimental spectra. In many cases, e.g. when only single particle observables are considered, this procedure is sufficient to phenomenologically describe a variety of aspects of the hadronic spectra (see e.g. \cite{STAR:2005_v2}). With regard to charge-balance correlations one needs to additionally incorporate local charge conservation. This can be achieved in the following way: Instead of generating a single particle at a time, we generate an ensemble of particles with exactly conserved charges, for a given ensemble every particle is then assigned a collective velocity $v_i$ such that all the $v_i$ follow the single particle blast-wave parameterization with the additional constraint of being emitted within a certain distance of each others. In the limit where charge conservation at kinetic freeze-out is perfectly local, this distance equals zero and all particles within a given ensemble are emitted as if their sources had the same collective velocity. If in contrast the balancing partner charges were spread out over the entire volume of the system, the locality constraint drops out and one recovers the conventional single particle blast-wave model. This extended freeze-out model then consists of three separate components, a parameterization of the chemical freeze-out configuration with exact charge conservation, a parameterization of the single particle freeze-out properties and a parameterization of the separation of the particles within an ensemble. \\
For the chemical composition we consider canonical ensembles at a chemical freeze-out temperature of $T_{chem}=175$ MeV \cite{PBM:2001_ChemicalFreezeout}  with total electric charge, strangeness and baryon number all equal to zero. We use a dilute description of the hadron resonance gas to calculate the canonical partition functions according to the procedure outlined in \cite{Cheng:2004_balance_models}. Finally we generate ensembles with zero net charge according to the canonical partition function. We consider mesons from the flavor octet and singlet ground state pseudoscalars and pseudovectors. Baryons are chosen from the ground state decuplet and octet. \\
For the single particle freeze-out properties we use the STAR parameterization suggested in \cite{STAR:2005_v2} that was shown to reproduce transverse momentum spectra as well as elliptic flow measurements. The transverse boost profile is given by an elliptic filled shell blast-wave, while the parameterization is boost invariant in the longitudinal direction. The parameters of the model are the kinetic freeze-out temperature $T_{kin}$, the maximum transverse collective rapidities in-plane $y_x$ and out-of-plane $y_y$ and the in-plane and out-of-plane radii $R_x$ and $R_y$ of the freeze-out surface. The points of emission $(x,y)$ are chosen uniformly within an ellipse with radii $R_x$ and $R_y$. For a given point $(x,y)$ on the freeze-out surface, the collective transverse rapidity is given by
\begin{eqnarray}
y_t&=&\tilde{r}~(\rho_0+\rho_2 \cos(2 \phi_B)) \; , \nonumber \\
\tilde{r}&\equiv&\sqrt{\left(\frac{x}{R_x}\right)^2 + \left(\frac{y}{R_y}\right)^2} \;, \nonumber \\
\mbox{atan}(\phi_B)&\equiv&\left(\frac{R_x}{R_y}\right)^2 \mbox{atan}\left(\frac{y}{x}\right) \;,
\label{eq:BlastWave}
\end{eqnarray}
where we introduced $\rho_0=(y_x+y_z)/2$ and $\rho_2=(y_x-y_z)/2$ according to the notation in \cite{STAR:2005_v2}. The direction of the transverse collective velocity is perpendicular to the surface of the ellipse at the point $(x,y)$, i.e. the collective velocity is given by $u^{\mu}=(\cosh(y_t),\sinh(y_t) \cos(\phi_B),\sinh(y_t) \sin(\phi_B),0)$. The situation is illustrated in Fig. \ref{fig:BlastWave}. We note that the parameterization is only sensitive to the ratio of the in-plane to the out-of-plane radius $R_x/R_y$ which has been extracted in \cite{STAR:2005_v2}, however the absolute values become meaningful when final-state interactions are considered. Hence this is addressed separately in Sec. III.a).\\

To incorporate local charge conservation one has to generate collective velocities for all particles of an ensemble in such a way that they satisfy a specified relative distribution, while the properties of the single particle distributions remain unchanged. For the relative distribution of the resonances within an ensemble we assume a gaussian distribution of the form
\begin{eqnarray}
\exp \left( -\frac{\Delta y_{z,ij}^2}{2 \sigma_\eta^2} \right) \exp \left( - \frac{\Delta\tilde{\phi_{ij}}^2}{2 \sigma_{\phi}^2}\right)
\label{eq:Gauss}
\end{eqnarray}
where $\tilde{\phi_i}=\mbox{atan}\left[(R_x/R_y)~(y_i/x_i)\right]$ as shown in Fig. \ref{fig:BlastWave}. Here $\Delta y_{z,ij}$ denotes the difference in longitudinal rapidity between the particles i and j, and accordingly for $\Delta\tilde{\phi_{ij}}$. This particular choice has two important properties that greatly simplify the sampling: a) The single particle distributions $P(y_{z,i})$ and $P(\tilde{\phi}_i)$ are uniform in $y_z$ (longitudinal boost invariance) and $\tilde{\phi}$ respectively. b) The use of a boost invariant blast-wave parameterization in combination with a relative distribution that factorizes into a longitudinal and a transverse part allows us to generate the longitudinal and transverse collective velocities separately.\\
For a given ensemble of $N$ particles the collective velocities can then efficiently be generated by a Metropolis algorithm, with a target distribution
\begin{eqnarray}
\int da \prod_{i=1}^N \left[P(y_{z,i})\right]^{1/N} \exp\left(-\frac{(y_{z,i}-a)^2}{\sigma_\eta^2}\right) \;,
\label{eq:Metropolis}
\end{eqnarray}
and accordingly for $\tilde{\phi}$. It is straightforward to verify that this reproduces the correct single particle distributions in the limit of $\sigma_\eta \frac{dP}{dy} \ll 1$. In practice we verify empirically that this is the case for all calculations presented in the upcoming sections by checking transverse momentum and elliptic flow. The integration over $a$ in Eq. (\ref{eq:Metropolis}) is part of the Metropolis routine, where the variable $a$ effectively represents the average $y_{z,i}$ and is introduced to accelerate convergence. \\
Finally the resonances and particles are assigned their thermal momenta according to the kinetic freeze-out temperature and their respective collective velocity. Resonances are then decayed according to measured branching ratios and lifetimes \cite{PDG:2002}. Since charge conservation is enforced on an ensemble-by-ensemble basis, particles originating from different ensembles are uncorrelated. Charge balance correlations can therefore efficiently be calculated on an ensemble-by-ensemble basis.

\section{III. SEPARATION OF BALANCING CHARGES}
\begin{figure}[htp]
\centerline{
	\includegraphics[width=0.45\textwidth]{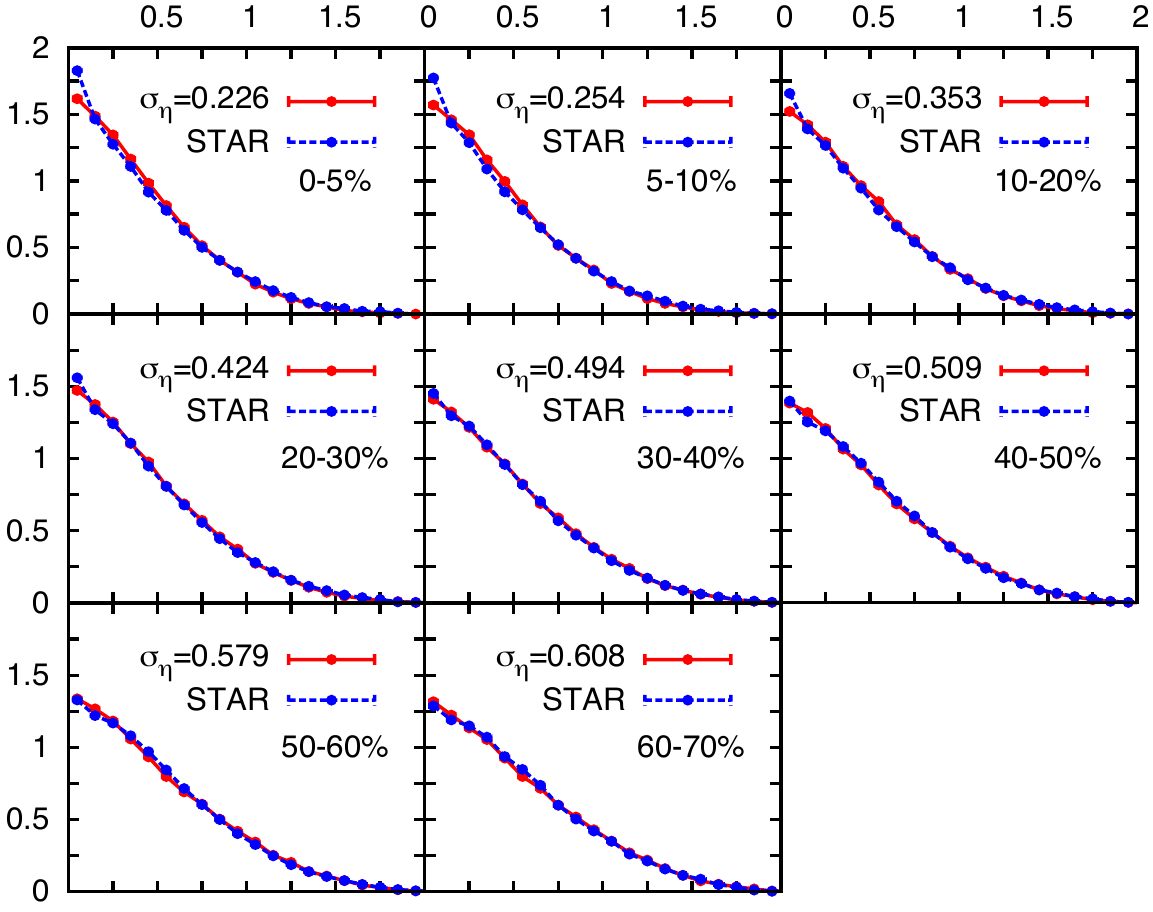}
}
\caption[Balance functions in relative pseudorapidity]%
	{\label{fig:BFdEta}
	Balance Function in $\Delta \eta$ for different centralities. The STAR data is shown in blue lines, the red curve is obtained from the modified blast-wave model (see Sec. II). The balancing charge separation in longitudinal rapidity $\sigma_\eta$ is indicated for each centrality bin. The broadening of the balance functions for less central collisions is a result of both the higher kinetic freeze-out temperature and a larger spatial separation of balancing charges at freeze-out.
}
\end{figure}

\begin{figure}[htp]
\centerline{
	\includegraphics[width=0.45\textwidth]{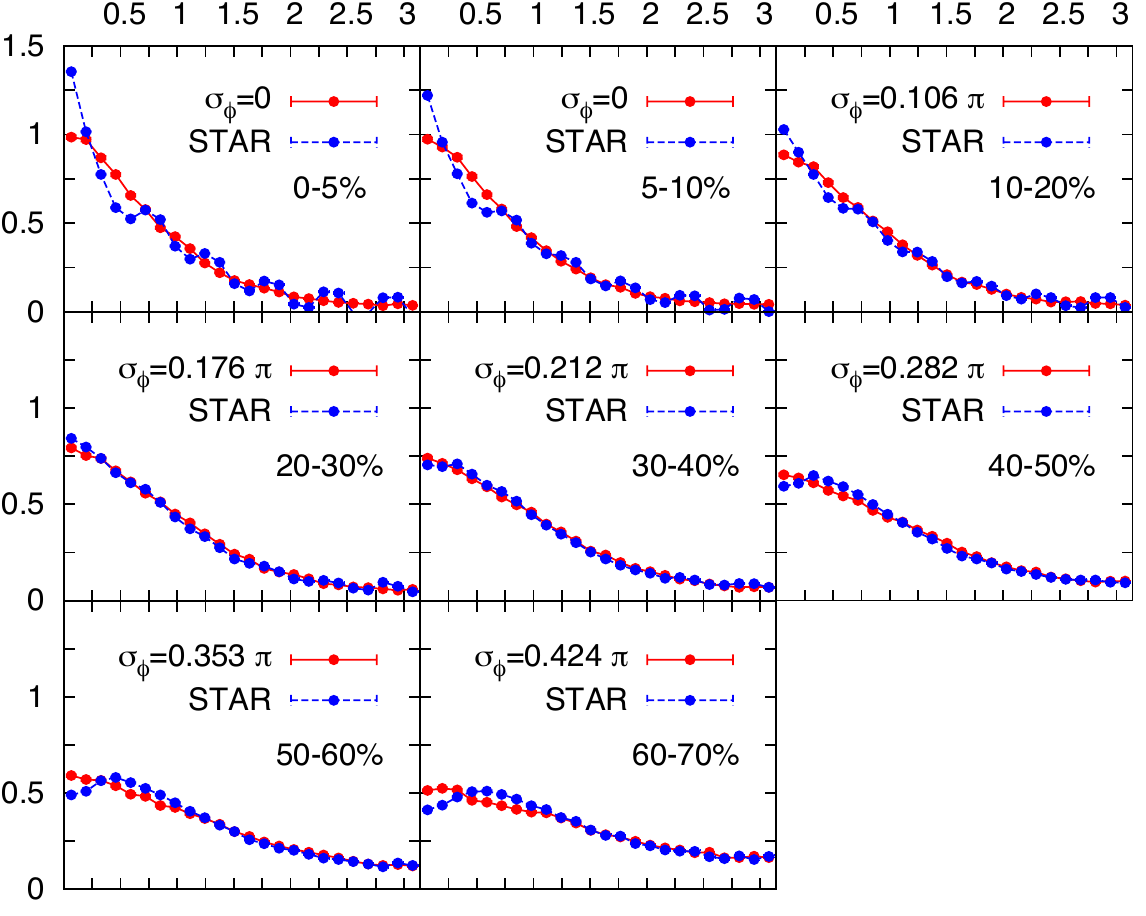}
}
\caption[Balance functions in relative azimuthal angle]%
	{\label{fig:BFdPhi}
	Balance Function in $\Delta \phi$ for different centralities. The STAR data is shown in blue lines, the red curve is obtained from the modified blast-wave model (see Sec. II). The balancing charge separation in azimuthal angle $\sigma_\phi$ is indicated for each centrality bin. The broadening of the balance functions for less central collisions is a result of the higher kinetic freeze-out temperature, less transverse flow and a larger separation of balancing charges at freeze-out.
}
\end{figure}

The separation of balancing charges at freeze-out is extracted from the experimental results \cite{STAR:2010_BalanceFunctions} in the following way. We calculate balance functions for all charged particles in relative pseudorapidity $\Delta \eta$ and relative azimuthal angle $\Delta \phi$ with the freeze-out model described in the previous section. We then adjust the widths of the relative distributions $\sigma_\eta$ and $\sigma_\phi$ for each centrality bin, in order to match the STAR results.  The single-particle blast-wave parameters  are taken from STAR \cite{STAR:2005_v2}, separately for each centrality. We assume perfect detector efficiency for the blast-wave calculation and use acceptance cuts of $0.2~\mbox{GeV/c}<p_t<2.0~\mbox{GeV/c}$ and $|\eta|<1.0$ as in the STAR measurement \cite{STAR:2010_BalanceFunctions}. The necessary efficiency correction is performed by normalizing all balance functions in such a way that they integrate to unity. In Fig. \ref{fig:BFdEta} we present the balance function in $\Delta \eta$ compared to the STAR data for various centralities. Here 0\% centrality corresponds to zero impact parameter and we refer to \cite{STAR:2010_BalanceFunctions} for more details on the classification. The considered charge separations are indicated in the figure and corresponds to respective best $\chi^2$-fit. Fig. \ref{fig:BFdPhi} shows the same results for balance functions in $\Delta \phi$.

\begin{figure}[htp]
\centerline{
	\includegraphics[width=0.45\textwidth]{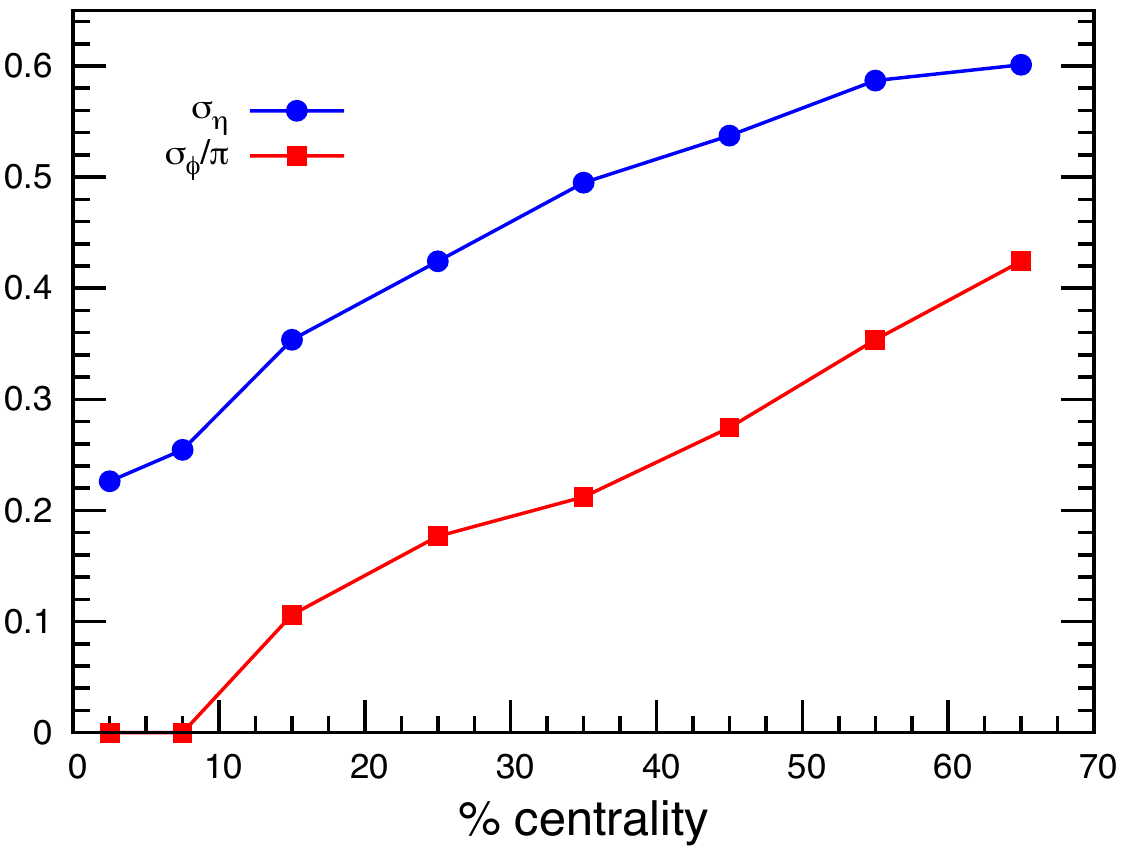}
}
\caption[Separation of balancing charges at freeze-out]%
	{\label{fig:separation}
	Separation of balancing charges in longitudinal rapidity $\sigma_\eta$ and azimuthal angle $\sigma_\phi$ at freeze-out as a function vs. centrality of the collision. The balancing charges are least separated both longitudinally and transversely for central collisions.
}
\end{figure}

The charge-balance correlations in $\Delta \eta$ show convincing agreement with the STAR data. We find that the observed broadening of the balance function for more peripheral collisions is a result of the higher kinetic freeze-out temperature and a larger separation of the balancing charges. For more central collisions the STAR data show an additional peak at small relative pseudorapidities which is not explained by the model. The overall behavior is nevertheless very well described by only considering charge-balance correlations. Also the consideration of additional sources of correlations has no significant effect on the result inferred for the longitudinal charge separation at freeze-out.\\
The agreement with the data is worse with regard to balance functions in $\Delta \phi$. First of all the STAR data shows oscillatory behavior for the three most central bins. This is expected to be an artifact caused by sector boundaries of the detector as discussed in great detail in the experimental analysis \cite{STAR:2010_BalanceFunctions}. However this issue complicates a reliable extraction of the charge separation. We note that the behavior at large $\Delta \phi$ is nevertheless well reproduced, even in this centrality region. Similar to the balance functions in $\Delta \eta$ we find that the STAR data shows an additional peak at small relative angles for very central collisions again not predicted by the model. In addition there is a bump in the experimental data at small relative angles for very peripheral collisions. We will show in Sec. III.a) how this arises from the distortion of the balance function when correlations due to final-state interactions are taken into account.\\
We emphasize that the change in the kinetic freeze-out temperature and the collective flow alone fail to explain the observed narrowing of the balance function for more central collisions. This feature of the STAR data can only be explained when the separation of balancing charges is taken into account. In Fig. \ref{fig:separation} we present our result on the separation of balancing charges at freeze-out. The results are obtained from the fit to charge balance functions. The separation of balancing charges in longitudinal rapidity is smallest for central collisions and increases for higher impact parameters. The qualitative behavior for the separation of balancing charges in azimuthal angle is similar, however there is a bigger uncertainty in the results primarily arising from the ambiguities with the experimental data. We restrain from showing error bars, as they could only include statistical errors and discuss the systematic uncertainties associated with the calculation instead. \\
The longitudinal results are fairly model independent, as the dependence on the parameterization in the transverse plane is weak and longitudinal boost invariance at mid-rapidity (recall that the analyzed data is obtained for $|\eta|<1$) is confirmed by various experimental observations. In contrast we expect a significant model dependence of the results in azimuthal angle, as the employed parameterization (see Eq. (\ref{eq:Gauss})) is somewhat arbitrary. Further insight with regard to this question might be achieved by the experimental study of more differential observables such as two dimensional charge balance functions in $\phi$ and $\Delta\phi$.  Finally there is some uncertainty associated with the chemical composition and the proper treatment of decays, in particular related to DCA cuts and detector efficiency, which in principle can be addressed by incorporating a more detailed efficiency and acceptance model into the calculation. We expect those issues to affect our inferences of $\sigma_\phi$ and $\sigma_\eta$ on the order of 10-20\% percent.

\subsection{a) INFLUENCE OF ADDITIONAL SOURCES OF CORRELATIONS}
\begin{figure}[ht]
\centerline{
	\includegraphics[width=0.45\textwidth]{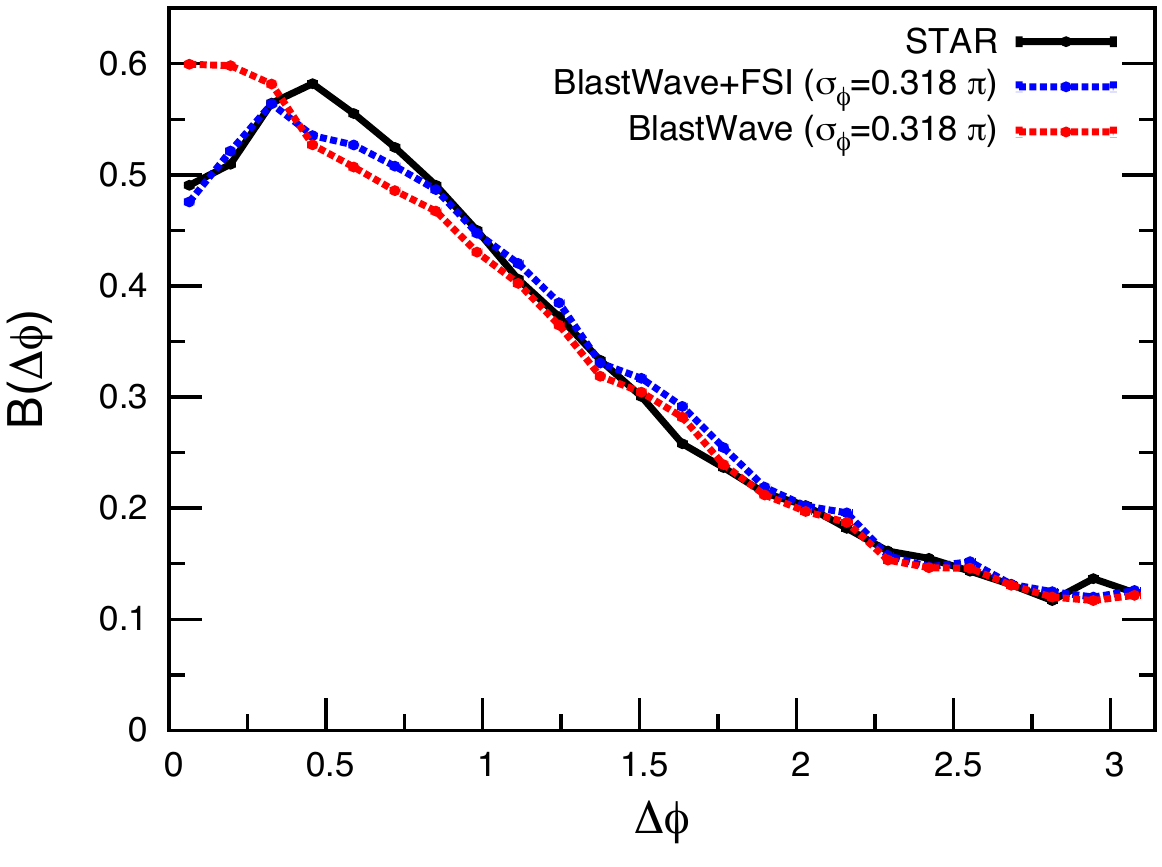}
}
\caption[Distortion of the balance function due to final-state interactions]%
	{\label{fig:distortion}
	Balance function $B(\Delta\phi)$ for 50-60$\%$ centrality from the blast-wave model (red curve). The corrections due to final-state interactions are included in the blue curve, STAR data is shown in black. The distortion of the balance function due to symmetrization and Coulomb interactions explains the behavior of the STAR data at small relative angles.
}
\end{figure}

In the previous discussion of balance functions it has been assumed that the balancing partner charge is statistically isolated by the like-sign subtraction $N_{+-}-N_{++}$, i.e. only the balancing partner contributes to this expression. This assumption does not hold when additional sources of correlations are taken into account and we have seen from the results presented in Fig. \ref{fig:BFdPhi}, that local charge conservation alone fails to explain the shape of the balance function for more peripheral collisions. Here we will consider final-state interactions between pion pairs to show that much of the experimentally observed discrepancy seen in Fig. \ref{fig:BFdPhi} is a result of final-state interactions.
In order to estimate the distortion of the balance function from residual interactions we follow the method provided in \cite{Cheng:2004_balance_models}. For every pair of balancing particles, one has to consider the correlation with other pairs of balancing particles. The correlation weight for a $\pi^+\pi^-$ pair, $p_a$ and $p_b$, with another $\pi^+\pi^-$ pair, $p_c$ and $p_d$, is then
\begin{eqnarray}
w(p_a,p_b;p_c,p_d)&=&|\psi_{ac}(p_a,p_c)|^2|\psi_{ad}(p_a,p_d)|^2 \nonumber \\ &&|\psi_{bc}(p_b,p_c)|^2|\psi_{bd}(p_b,p_d)|^2 
\label{eq:corr_weight}
\end{eqnarray}
where $\psi_{\alpha\beta}$ is the Coulomb wavefunction with asymptotic momentum $p_\alpha-p_\beta$. For two pions of the same sign, the wavefunctions are symmetrized. The distortion of the balance function due to identical particle symmetrization and Coulomb correlations is then estimated from the additional contribution to the balance function numerator from the two pairs, where instead of increasing the pair distributions $N_{++}$ and $N_{+-}$ in unit steps, the distributions are increased by the respective correlation weight $w(p_a,p_b;p_c,p_d)$. The pairs are generated according to the blast-wave prescription described in Sec. II. We restrict to $\pi^+\pi^-$ pairs and assume that the two particles from the same pair are emitted from the same collective velocity. The charged particle multiplicities are chosen according to STAR measurements \cite{STAR:2009_Mult} and we choose a source radius of $\sqrt{R_x^2+R_y^2}=13~\mbox{fm}$ in the transverse plane. The longitudinal size of the system is determined by the freeze-out time which is chosen to be $\tau=10~\mbox{fm/c}$ to reproduce HBT spectra \cite{Cheng:2004_balance_models}. The distortion of the balance is then estimated by applying the corrections to 25\% of all pairs, in order to account for pions emitted from long-lived decays and because only a fraction of pairs involve two pions. This estimate is rather crude, but should be sufficient to see if the effects of final-state interactions are of the appropriate magnitude and shape to describe the discrepancies seen in Fig. \ref{fig:BFdPhi}.\\
In Fig. \ref{fig:distortion} we show how the balance function for 50-60$\%$ centrality is distorted due to symmetrization and Coulomb correlations of $\pi^+\pi^-$ pairs. We find that the dip at small $\Delta\phi$ present in the STAR data, can be reproduced when identical particle symmetrization and Coulomb correlations are taken into account. However this happens at the expense of higher values of the balance function at large $\Delta\phi$, and therefore also affects the separation of balancing charges extracted in the previous section.  In particular for the 50-60$\%$ centrality bin we find that the distortion of the balance function lowers $\sigma_\phi$ by about ten percent. The effects of additional sources of correlations on the balance function in $\Delta\eta$ are expected to be significantly lower. This is because pairs of particles with small $\Delta\phi$ are also within two units of pseudorapidity, whereas for a pair of particles with small $\Delta\eta$ they the particles can be spread out over the entire angle of $2\pi$. Hence one would naturally expect the effects to be smaller by a factor of $\pi$. For a more detailed study of the residual sources of correlations non-resonant contributions of the strong interaction also have to be taken into account and the analysis has to be extended to other particle species. We recommend performing such studies in terms of balance functions of the invariant momentum of the pair $q_{inv}$ and its components $q_{out}$, $q_{side}$ and $q_{long}$. 

\section{IV. CONTRIBUTIONS TO LOCAL PARITY VIOLATION OBSERVABLES}
\begin{figure}[hpt]
\centerline{\includegraphics[width=0.45\textwidth]{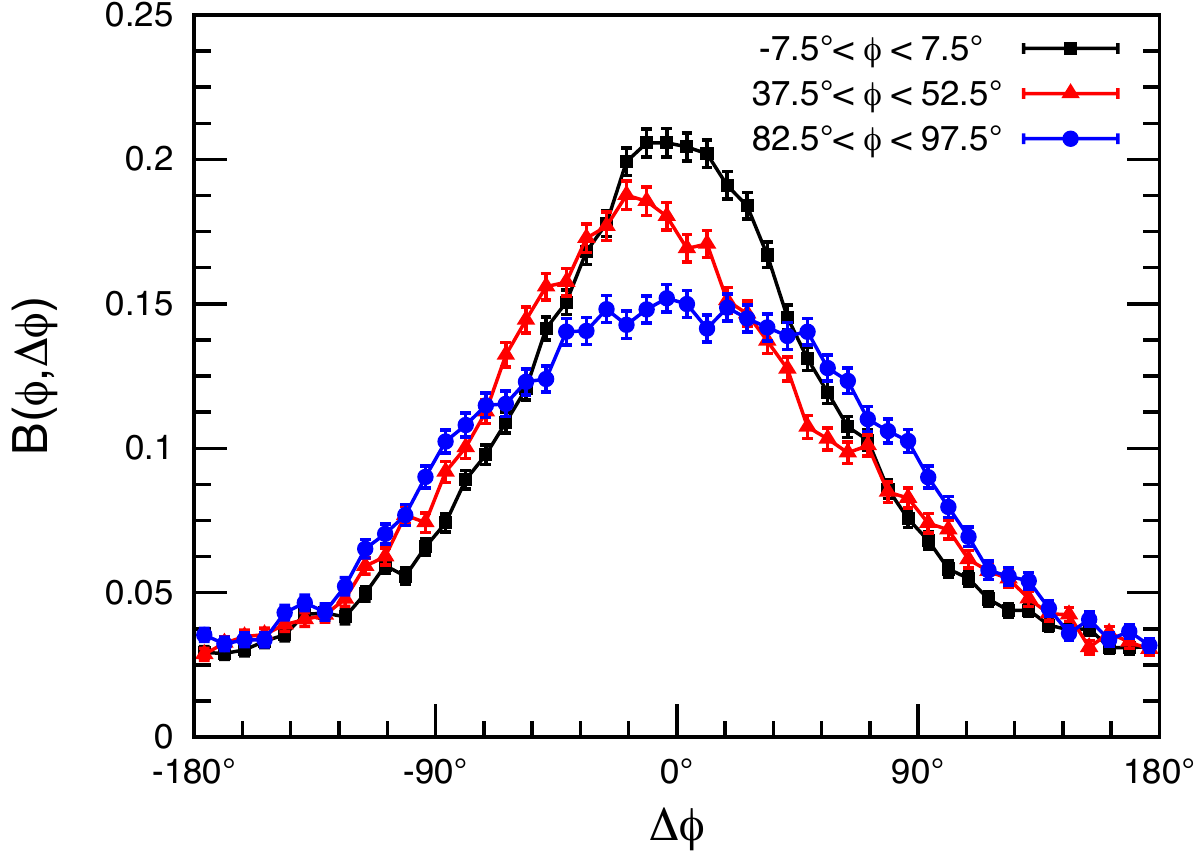}}
\caption[Reaction plane dependent balance functions]
	{\label{fig:balance}
	Balance function $B(\phi,\Delta\phi)$ for 40-50\% centrality shown as function of the relative angle included by balancing partners for $\phi=0^\circ$ (black squares), $45^\circ$ (red triangles) and $90^\circ$ (blue circles). The balance function is narrower for in-plane pairs compared to out-of-plane pairs. For intermediate angles the balance function is biased toward negative angles.}
\end{figure}

The prospect of observing a signature of topologically non-trivial QCD gauge field configurations in relativistic heavy ion collisions has recently gained great attention \cite{STAR:2009_Parity,Voloshin:2004}. It was shown that in the presence of such configurations $\mathcal{P}$- and $\mathcal{CP}$- violating processes occur locally in regions with non-zero QCD topological charge \cite{Kharzeev:2009_Parity}, which are expected to be created in the early stage of relativistic heavy ion collisions. In particular in the presence of a magnetic field positive charge is separated from negative charge along the magnetic field \cite{Kharzeev:2010_Parity}. This phenomenon is referred to as the ``Chiral magnetic effect''. In the situation of relativistic heavy ion collisions the magnetic field is induced by the colliding nuclei. This implies that parity is locally violated on an event by event basis as positive and negative charges separate along the magnetic field, which is perpendicular to the reaction plane (out-of-plane) \cite{STAR:2009_Parity}. However because there is no direct $\mathcal{P}$- and $\mathcal{CP}$- violation in QCD, the probability to generate charge separation parallel or anti-parallel to the magnetic field is equal and the expectation value of any $\mathcal{P}$-odd observable vanishes. The possibilities to identify the existence of a local parity violating effect are therefore confined to the study of fluctuations of $\mathcal{P}$-odd observables (which are then $\mathcal{P}$-even). The observable measured by STAR is
\begin{equation}
\gamma_{\alpha,\beta}=\frac{\sum_{i \in \alpha,j \in \beta} \cos(\phi_i+\phi_j -2 \psi_{RP})}{M_\alpha M_\beta},
\label{eq:STAR_PO}
\end{equation}
where $\alpha$ and $\beta$ represent positive or negative charge, $M_\alpha$ and $M_\beta$ are the corresponding multiplicities, the azimuthal angles $\phi$ are measured about the beam axis and $\psi_{RP}$ is the angle of the reaction plane which we will set to zero in the following without loss of generality. The observable is $\mathcal{P}$-even, so that in principle other sources of correlations can contribute to the signal. Since one is looking for charge separation along the out-of-plane direction, the idea is to compare the charged particle correlations out-of-plane to the corresponding in-plane-correlations. By use of the identity $\cos(\phi_i+\phi_j)=\cos(\phi_i)\cos(\phi_j)-\sin(\phi_i)\sin(\phi_j)$, the STAR observable subtracts in-plane from out-of-plane correlations to eliminate other sources of correlation that are independent of the orientation w.r.t the reaction plane. If there are no reaction-plane-dependent background contributions, charge separation in out-of-plane direction then causes $\gamma$ to be negative for same-sign pairs and positive for opposite-sign pairs. The correlators in Eq. (\ref{eq:STAR_PO}) have recently been measured by the STAR collaboration \cite{STAR:2009_Parity}. In addition to the integrated signal differential measurements have been performed in relative pseudorapidity $\Delta \eta$ and transverse momentum $p_t$ \cite{STAR:2009_Parity}. The reported results are qualitatively in line with various expectations from the chiral magnetic effect and significantly larger than reaction-plane-dependent background contributions from MEVSIM, HIJING and UrQMD simulations \cite{STAR:2009_Parity}. However there is an ongoing discussion on the interpretation of the data, in particular if the chiral magnetic effect can produce a signal of the observed order of magnitude \cite{Asakawa:2010_Parity,Pratt:2010_Parity} and whether there are more ``traditional'' explanations of the observed signal (e.g. cluster particle correlations \cite{FWang:2010}; charge conservation and flow \cite{Schlichting:2010}, momentum conservation \cite{Pratt:2010_Parity,Bzdak:2010_2}). Furthermore the separate analysis of in-plane and out-of-plane correlations has revealed, that the correlations for same-sign particles are mainly in-plane and back-to-back \cite{Bzdak:2010}, indicating the presence of other sources of correlations, and additional measurements have been proposed to clarify the situation \cite{Voloshin:2010}. In the following we will show how charge balance gives rise to reaction-plane-dependent correlations and calculate the respective contributions to the local parity violation observable from the blast-wave model described in Sec. II.\\
From the previous discussion (see Sec. I) it is clear that charge-balance correlations only contribute to the difference of opposite-sign and same-sign pair correlations. We will therefore only discuss contributions to
\begin{eqnarray}
\label{eq:gammapdef}
\gamma_P&\equiv&\frac{1}{2}\left(2\gamma_{+-}-\gamma_{++}-\gamma_{--}\right).
\end{eqnarray}
The observable $\gamma_P$ then compares in-plane vs. out-plane correlations for opposite-sign vs. same-sign charged pairs. In this context the discussion of charge-balance correlations has to be extended to balance functions sensitive to the angle of the pair with respect to the reaction plane $\phi$, i.e. the object of interest is
\begin{eqnarray}
\label{eq:bf_p}
B(\phi,\Delta \phi)&=&\frac{1}{dM/d\phi}\int dp_1 \frac{dM}{dp_1} dp_2~B(p_1,p_2) \\ &&\delta(\phi-\phi_1)\delta(\Delta\phi-(\phi_2-\phi_1)) \nonumber \;.
\end{eqnarray}
Balance functions $B(\phi,\Delta\phi)$ are presented as a function of $\Delta\phi$ in Fig. \ref{fig:balance} for events with centralities of 40-50\%. The results are obtained from the blast-wave model described in Sec. II for the charge-separation parameters extracted in Sec. III. The balance function for $\phi=0^\circ$ (in-plane) is narrower than the balance function for $\phi=90^\circ$ (out-of-plane). The stronger focusing of balancing charges derives from the greater collective flow in-plane vs. out-of-plane. For $\phi=45^\circ$, the distribution is biased toward negative values of $\Delta\phi$. This is expected given the elliptic asymmetry, $v_2>0$, which leads to more balancing particles toward the $\phi=0^\circ$ direction as opposed to $\phi=90^\circ$. Depending on which quadrant $\phi$ is located, the balancing charge tends to be found more toward $\phi=0^\circ$ or $\phi=180^\circ$.

\begin{figure}[hpt]
\centerline{\includegraphics[width=0.45\textwidth]{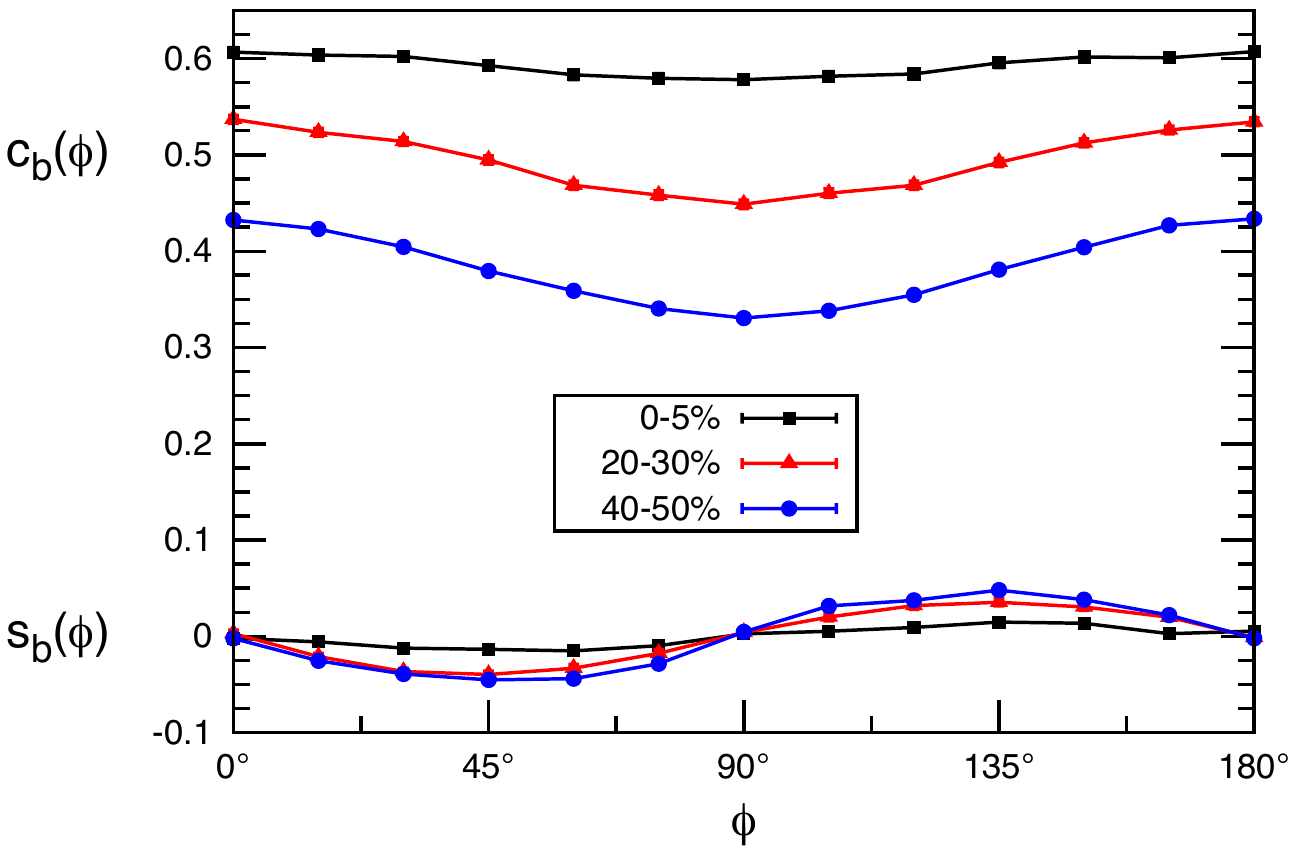}}
\caption[Moments of the balance function]
	{\label{fig:moments}The moments of the balance function, $c_b(\phi)$ and $s_b(\phi)$, represent averages of $\cos(\Delta\phi)$ and $\sin(\Delta\phi)$ across the balance function. These are plotted as a function of $\phi$ for various centralities. The structure of $c_b(\phi)$, which is maximized at $\phi=0^\circ,180^\circ$, illustrates how the balance function is narrower for in-plane emission and for more central collisions, while the structure of $s_b(\phi)$, which is positive for $\phi=135^\circ,315^\circ$ and negative for $\phi=45^\circ,225^\circ$, shows how balancing charges prefer to be emitted in the in-plane direction. The oscillations increase for more peripheral collisions.
}
\end{figure}

The width and asymmetry of the balance function can be quantified by the moments, 
\begin{eqnarray}
\label{eq:bfmoments}
c_b(\phi)&\equiv&\frac{1}{z_b(\phi)}\int d\Delta\phi~B(\phi,\Delta\phi)\cos(\Delta\phi),\\
\nonumber
s_b(\phi)&\equiv&\frac{1}{z_b(\phi)}\int d\Delta\phi~B(\phi,\Delta\phi)\sin(\Delta\phi),
\end{eqnarray}
where 
\begin{eqnarray}
z_b(\phi)&\equiv&\int d\Delta\phi~B(\phi,\Delta\phi),
\end{eqnarray}
is the normalization of the balance function and represents the probability of detecting the balancing charge given the observation of a charge at $\phi$. It would be unity for a perfect detector, but is reduced by both the finite acceptance and efficiency of the experiment. The moments $c_b(\phi)$ and $s_b(\phi)$ are displayed in Fig. \ref{fig:moments}. The quantity $c_b(\phi)$ expresses the width of the balance function and would be unity for a very narrow balance function whereas it vanishes in the case where the balancing charges were emitted randomly. The quantity $s_b(\phi)$ measures the degree to which the balance function is asymmetric under a reflection symmetry of $\Delta\phi \rightarrow -\Delta\phi$. For pairs around $\phi=45^\circ$ this corresponds to the probability for the balancing charge to be emitted in in-plane direction vs. in out-of-plane direction.\\

\begin{figure}[ht]
\centerline{
	\includegraphics[width=0.45\textwidth]{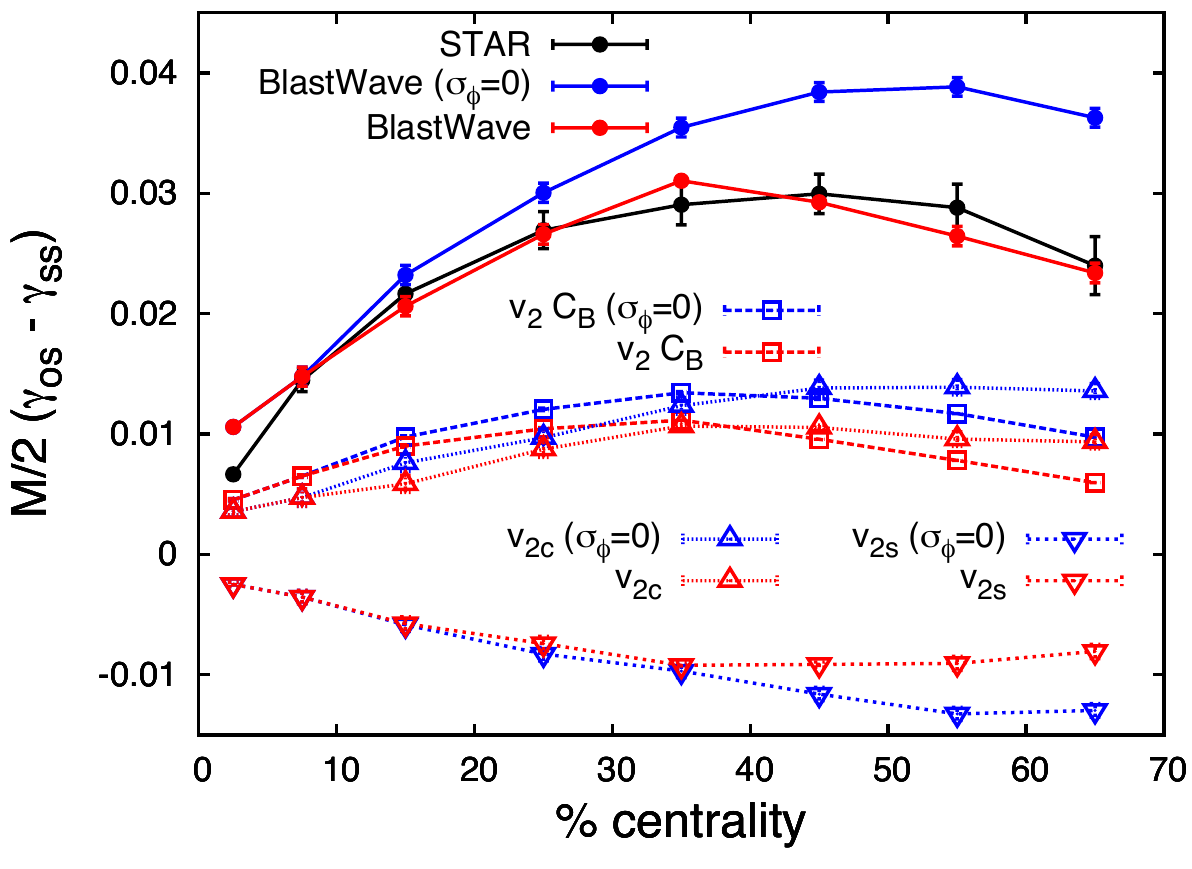}
}
\caption[Charge balance contributions to STAR's local parity violation observable]%
	{\label{fig:gamma}
	Difference between opposite-sign and same-sign parity observable from STAR (black dots) and blast-wave calculations for realistic charge separation at freeze-out (red dots) and perfectly local charge conservation (blue dots). The three contributions to the signal are defined in Eq. (\ref{eq:components}) and are plotted with dashed lines. $v_2\langle c_b\rangle$ (squares) derives from having more balancing pairs in-plane than out-of-plane while $v_{2,c}$ (triangles up) quantifies the degree to which in-plane pairs are more tightly correlated than out-of-plane pairs. $v_{2,s}$ (triangles down) reflects that the balancing charge is more likely to be found toward the event plane. 
}
\end{figure}

In order to relate the correlation $\gamma_P$ to moments of the balance function $B(\phi,\Delta\phi)$, some remarks on the event averaging are in order. While the event average of $\gamma_P$ is simply given by $\langle \gamma_P \rangle$ \cite{STAR:2009_Parity}; charge balance functions are calculated as the ratio of event averages,
\begin{eqnarray}
\label{eq:bf_avg}
 B(\phi,\Delta\phi)&=&\frac{\langle N_{+-}(\phi|\Delta\phi)-N_{++}(\phi|\Delta\phi) \rangle}{\langle dM/d\phi \rangle}\\
&&+\frac{ \langle N_{-+}(\phi|\Delta\phi)-N_{--}(\phi|\Delta\phi)\rangle}{\langle dM/d\phi \rangle} \;. \nonumber
\end{eqnarray}
By use of the angle addition formula $\cos(\phi_i+\phi_j)=\cos(2\phi_i)\cos(\Delta\phi)-\sin(2\phi_i)\sin(\Delta\phi)$ we find the expression relating the correlation $\gamma_P$ to moments of the balance function,
\begin{eqnarray}
\label{eq:relation}
\frac{\langle M^2 \gamma_P \rangle}{\langle M \rangle}&=&\frac{2}{\langle M \rangle} \int d\phi~d\Delta\phi~\left\langle\frac{dM}{d\phi}\right\rangle \; B(\phi,\Delta\phi)\\
&&\qquad \quad \left[\cos(2\phi)\cos(\Delta\phi) -\sin(2\phi)\sin(\Delta\phi)\right], \nonumber
\end{eqnarray}
where it has been assumed that there are equal numbers of positive and negative charges and the factor of $1/\langle M \rangle$ makes the result independent of the multiplicity. In the limit of very small multiplicity bins, the left-hand side of Eq. \ref{eq:relation} could be replaced by $\langle M \rangle\langle\gamma_P\rangle$\\
It is insightful to express the correlations, $\gamma_P$ , in terms of the moments of the balance function defined in Eq. (\ref{eq:bfmoments}),
\begin{eqnarray}
\label{eq:components}
\frac{\langle M^2\gamma_P \rangle}{2\langle M \rangle}&=&v_2\langle c_b(\phi)\rangle+v_{2c}-v_{2s},
\end{eqnarray}
where we introduced
\begin{eqnarray}
v_{2c}&\equiv&\langle c_b(\phi)\cos(2\phi)\rangle-v_2\langle c_b(\phi)\rangle, \nonumber \\
v_{2s}&\equiv&\langle s_b(\phi)\sin(2\phi)\rangle,\nonumber \\
\langle f(\phi)\rangle &\equiv& \frac{1}{\langle M \rangle}\int d\phi~\frac{dM}{d\phi}~z_b(\phi) f(\phi).
\end{eqnarray}
The three contributions to $\gamma_P$ derive from: a) having more balancing pairs in-plane than out-of-plane $(v_2\langle c_b\rangle)$, b) having the in-plane pairs being more tightly correlated in $\Delta\phi$ than the out-of-plane pairs $(v_{2c})$ and c) having the balancing charge more likely being emitted toward the event plane $(v_{2s})$.\\
The three contributions to the signal obtained from the blast-wave calculation are displayed in Fig. \ref{fig:gamma}.  We assume perfect detector efficiency for the blast-wave calculation and use the same acceptance cuts in transverse momentum and pseudorapidity, i.e. $0.15~\mbox{GeV} < p_t < 2~\mbox{GeV}$ and $|\eta|<1$. The necessary efficiency correction is done by rescaling the results to reproduce the experimental normalization of the balance function \cite{STAR:2010_BalanceFunctions}, i.e. we multiply the expressions for $\langle c_b(\phi) \rangle$, $v_{2c}$ and $v_{2s}$ by the ratio of experimental to blast-wave normalization. In addition to the results obtained for the (realistic) charge separation extracted in Sec. III, we present the contributions assuming that charge conservation is perfectly local at freeze-out. These are the strongest possible contributions to the signal and should give an upper limit. In order to compare to STAR data \cite{STAR:2009_Parity} we modify the left hand side of (\ref{eq:components}) in the following way. Since $M \gamma_P$ is independent of the multiplicity we expect $\langle M^2 \gamma_P \rangle\approx\langle M \rangle \langle M\gamma_P\rangle$, with $<M\gamma_P>$ contained in $M\langle \gamma_P \rangle$ within errorbars. The left hand side of (\ref{eq:components}) then simplifies to $M/2 \langle \gamma_P \rangle$, where $M$ is the experimental multiplicity for a single event accounting for efficiency and acceptance of the detector \cite{STAR_mult}.

\begin{figure}[hpt]
\centerline{
	\includegraphics[width=0.45\textwidth]{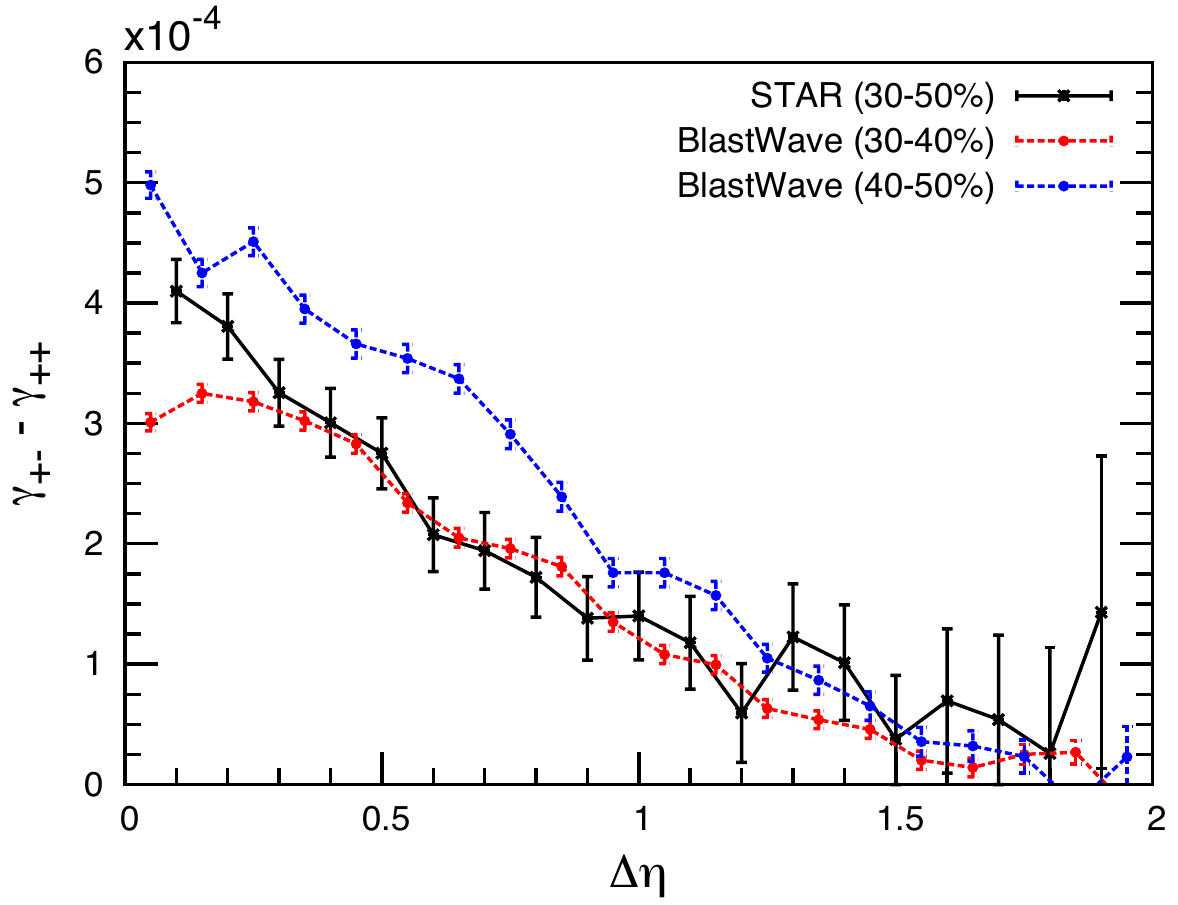}
}
\caption[Contributions to STAR's local parity violation observable differential in relative pseudorapidity]%
	{\label{fig:diff_eta}
	Differential Parity Observable from STAR for 30-50\% centrality (black solid) and blast-wave calculations for 30-40\% (red dashed) and 40-50\% centrality (blue dashed). The correlations decay in $\Delta\eta$ as balancing charges tend to be emitted in a narrow range of relative pseudorapidity.
}
\end{figure}

\begin{figure}[hpt]
\centerline{
	\includegraphics[width=0.45\textwidth]{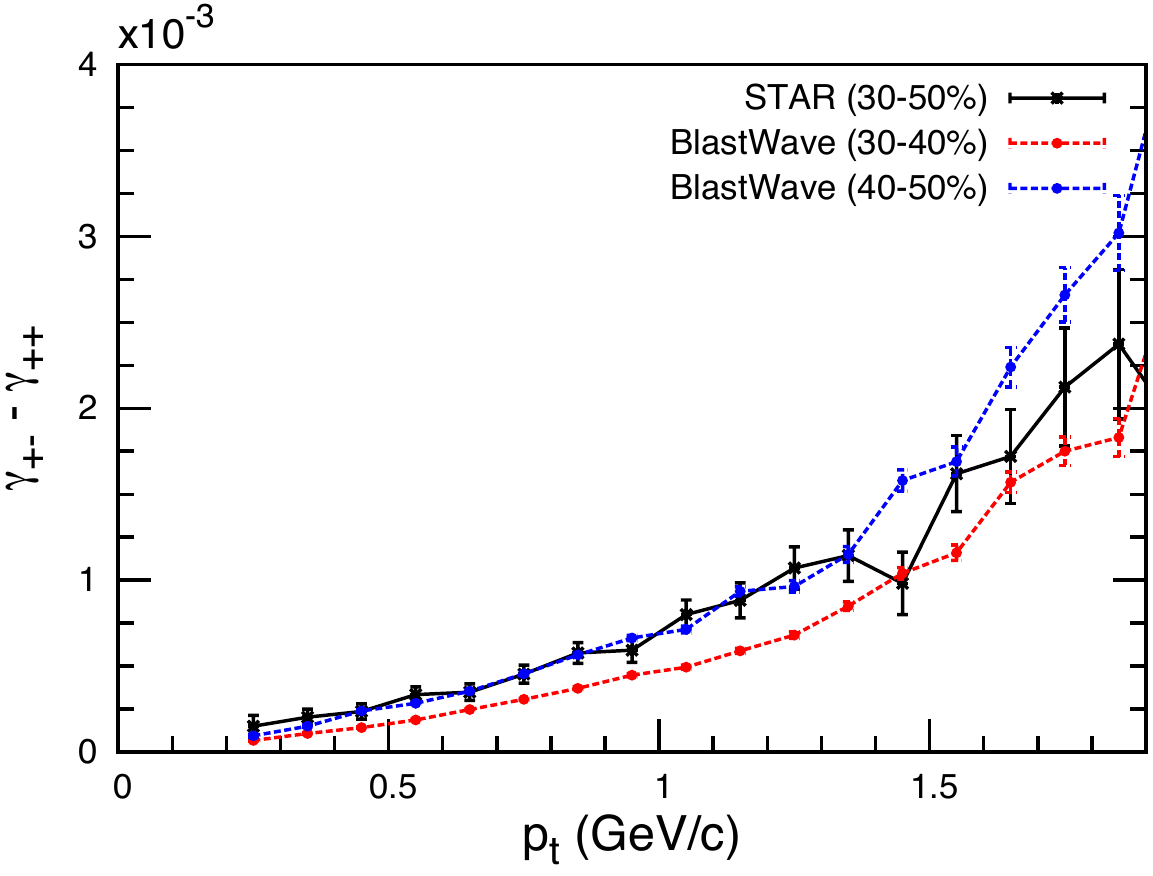}
}
\caption[Contributions to STAR's local parity violation observable differential in transverse momentum]%
	{\label{fig:diff_pt}
	Differential Parity Observable from STAR for 30-50\% centrality (black solid) and blast-wave calculations for 30-40\% (red dashed) and 40-50\% centrality (blue dashed). The correlations increase in the transverse momentum of the pair $p_t=(p_{t,i}+p_{t,j})/2$, because of higher anisotropy and more collective flow.
}
\end{figure}

The contributions of charge-balance correlations for a realistic charge separation are of the same size as the experimental signal and exhibit similar qualitative behavior with respect to the centrality dependence. The systematic error associated with this prediction originates predominantly from the particular parameterization of the charge separation in azimuthal angle (see Sec. II, Eq. (\ref{eq:Gauss}) for more details) and we expect the conclusions to be reliable on a ten percent accuracy level.\\
In the present discussion of the interpretation of STAR results it has been proposed to analyze the in-plane and out-of-plane correlations separately \cite{Bzdak:2010}. With respect to the difference between opposite-sign and same-sign correlations this becomes trivial as the sum of in-plane and out-of-plane correlations is related to reaction-plane-independent balance functions by
\begin{eqnarray}
&&\langle\cos(\phi_i-\phi_j)\rangle_{+-}-\langle\cos(\phi_i-\phi_j)\rangle_{++}= \nonumber \\
&&\qquad \int d\Delta\phi~B(\Delta\phi)~\cos(\Delta\phi) \;,
\end{eqnarray}
which are well understood from the physics of charge balance as discussed in detail in the previous section. The correlations can then be separated into their in-plane and out-of-plane components by the use of trigonometric identities \cite{Bzdak:2010}, both showing similar agreement with the STAR data \cite{STAR:2009_Parity} as the results presented in Fig. \ref{fig:gamma}.\\
In addition to the integrated signal the correlations have been analyzed differentially in pseudorapidity $\Delta\eta$ and transverse momentum $p_{t,i}+p_{t,j}$ \cite{STAR:2009_Parity}. In order to relate the differential correlators to moments of the balance function, one has to account for the number of pairs in each differential bin. Hence Eq. (\ref{eq:relation}) has to be modified to,
\begin{eqnarray}
\frac{\left\langle N_p(\Delta\eta)\gamma_P(\Delta\eta)\right\rangle}{\langle M \rangle}&=&\frac{2}{\langle M \rangle}\int d\phi~\left\langle\frac{dM}{d\phi}\right\rangle~d\Delta\phi \\
&&\qquad B(\phi,\Delta\phi,\Delta\eta)~\cos(2\phi+\Delta\phi) \nonumber \;,
\label{eq:diff_gamma}
\end{eqnarray}
where $N_p(\Delta\eta)/M^2$ is the fraction of charged particle pairs in the respective pseudorapidity bin. The right hand side of (\ref{eq:diff_gamma}) can be obtained straightforward from the blast-wave model. In order to compare to experimental data we will assume that, for a given event, the correlation $\gamma_P(\Delta\eta)$ scales with the total multiplicity rather than the fraction of pairs in the respective bin, i.e.
\begin{eqnarray}
\left\langle \frac{N_p(\Delta\eta)}{M^2} M^2 \gamma_P(\Delta\eta) \right\rangle 
 &\approx& \left\langle \frac{N_p(\Delta\eta)}{M^2} \right \rangle \langle M^2 \gamma_P(\Delta\eta) \rangle \nonumber \\
 &\approx& \left\langle \frac{N_p(\Delta\eta)}{M^2} \right \rangle M \langle M \rangle \langle \gamma_P(\Delta\eta) \rangle \nonumber \; ,
\end{eqnarray}
where the average fraction of pairs per pseudorapidity bin can be obtained from the single particle spectra of the blast-wave model. The charge balance contributions to the differential observables are presented in Fig. \ref{fig:diff_eta} and \ref{fig:diff_pt} respectively. The blast-wave results are obtained for realistic charge separation at freeze-out, the qualitative behavior is similar to what is observed at STAR. When regarded differentially in pseudorapidity the correlations are confined to a finite range in $\Delta\eta$. This is because balancing charges are likely to be emitted within a narrow range of rapidity. This can already be seen from balance functions in relative pseudorapidity (see Sec. II). When regarded differentially in $p_t$ the correlations increase with the momentum of the pair due to larger anisotropy (i.e. higher $v_2$) and more collective flow.

\section{V. CONCLUSIONS}
We have seen in Sec. III that the charge balance functions recently observed by the STAR collaboration \cite{STAR:2010_BalanceFunctions} can be understood as a signal of highly localized charge balance at kinetic freeze-out. The observed narrowing of the charge balance function for more central collisions, that could not be observed in URQMD and HIJING simulations \cite{STAR:2010_BalanceFunctions}, is explained by a smaller separation of balancing charges at freeze-out, which is not inherent to these models. The notion, that the observed narrowing of the balance function for more central collisions is entirely due to a reduced freeze-out temperature and increased transverse flow, can be dismissed. Even though a reduced freeze-out temperature and increased transverse flow lead to higher correlations of balancing charges in more central collisions, the relevant parameters describing the flow and temperature of the breakup stage are constrained by spectra and elliptic flow measurements.\\
As seen in Fig. \ref{fig:separation}, the separation of balancing charges in $\Delta\eta$ appears smaller for central collisions than for peripheral collisions. This separation is related to the separation in coordinate space and the production time through the relation $\sigma_\eta \approx \sigma_z/\tau$. In p+p collisions the separation is driven by the dynamics of breaking gluon-strings, or color flux tubes. By tunneling out of the vacuum, balancing charges typically appear separated by $\sim1/2~\mbox{fm}$ along the z-direction. If the particles are produced $\sim1/2~\mbox{fm/c}$ after the initial impact, they are already separated by a good fraction of a unit in pseudorapidity. This separation can only increase in the subsequent time evolution, as the charges diffuse relative to one another.\\
The observation of smaller $\sigma_\eta$ for central Au+Au collisions implies that either:
\begin{itemize}
 \item[a)] Charges are produced at later times, as this would allow the same $\sigma_z$ to give a smaller $\sigma_\eta$. Late charge production is expected from delayed hadronization, as the formation processes of hadrons from quarks and gluons involve the creation of additional charge.
\item[b)] Charges are produced early by a different mechanism than the decay of color flux tube. In this situation $\sigma_z$ would be smaller initially, while $\tau$ remains small. For instance, if the matter became isotropized gluonic matter at very early times, quark production might result from collisions rather than from tunneling through longitudinal flux tubes. This scenario also requires charge diffusion to be small, to constrain $\sigma_\eta$ at freeze-out close to the initial separation at creation.
\end{itemize}
Both of these explanations, or a combination of the two, might explain the smaller separation of balancing charges $\sigma_\eta$ for more central collisions. The results presented in Sec. III therefore point to a change in the charge-production mechanism from p+p to central Au+Au collisions. However to draw further conclusions it is necessary to compare the extracted separations of balancing charges to transport properties of the hadron resonance gas and the quark gluon plasma. As a first step it would be important to know how much the particles diffuse between chemical and kinetic freeze-out. These results could in principle be obtained from hadronic cascade models or transport theory.\\
The separation of balancing charges in azimuthal angle, described by $\sigma_\phi$, shows a similar behavior with respect to the centrality dependence. As the size of the system in the transverse plane increases with centrality, this is qualitatively expected as long as the charges can not diffuse far away from one another, which would require both early production time and large diffusion. Hence the analysis of charge separation in azimuthal angle does not give any further insight concerning charge production mechanisms.\\
With respect to STAR's local parity violation observable, we have seen that local charge conservation gives rise to a reaction-plane-dependent ``background'' signal, contributing to the difference of opposite-sign and same-sign correlations. The balancing charge correlations are of the same size as the observed correlations and exhibit similar qualitative behavior with respect to all measured properties. This suggests that local charge conservation is the dominant effect with regard to the difference of same-sign and opposite-sign parity observable. We emphasize that this does not explain the experimentally observed strong same-sign correlations. However, it is clear that neither charge balance or parity fluctuations can produce same-sign correlations without affecting the difference of opposite-sign and same-sign correlations. A mechanism that has been suggested to produce strong same-sign correlations is momentum conservation \cite{Pratt:2010_Parity}. This is the topic of a separate study.

\acknowledgments{This work was supported by the U.S. Department of Energy, Grant No. DE-FG02-03ER41259.}


\begin{thebibliography}{99}
\bibitem{STAR_mult} Experimental multiplicities were generously provided by the STAR collaboration.
\end{thebibliography}
\end{document}